\documentclass[10pt,prd,twocolumn,floatfix,preprintnumbers,nofootinbib,superscriptaddress,longbibliography]{revtex4-1}
\pdfoutput=1
\usepackage[colorlinks=true,breaklinks=true]{hyperref}
\hypersetup{allcolors=[rgb]{0.0 0.0 0.6},linkcolor=[rgb]{0.0 0.5 0.6}}
\usepackage{amsmath,amssymb}
\usepackage{epsfig}  
\usepackage{graphicx}   
\usepackage{slashed}             
\usepackage{url}
\usepackage{color}
\usepackage{multirow}
\usepackage{placeins}
\usepackage[dvipsnames]{xcolor}
\usepackage{letltxmacro}
\LetLtxMacro{\oldcite}{\cite}
\renewcommand{\cite}[1]{\mbox{\oldcite{#1}}}
\newcommand{\gag}{g_{a\gamma}}

\allowdisplaybreaks

\bibpunct{[}{]}{,}{n}{}{,}

\begin{document}

\preprint{LAPTH-042/20}

\title{Bounds on axionlike particles from the diffuse supernova flux}

\author{Francesca Calore}
 \email{calore@lapth.cnrs.fr}
\affiliation{Univ.~Grenoble Alpes, USMB, CNRS, LAPTh, F-74000 Annecy, France}

\author{Pierluca Carenza}
\email{pierluca.carenza@ba.infn.it }
\affiliation{Dipartimento Interateneo di Fisica ``Michelangelo Merlin'', Via Amendola 173, 70126 Bari, Italy.}
\affiliation{Istituto Nazionale di Fisica Nucleare - Sezione di Bari,
Via Orabona 4, 70126 Bari, Italy.}

\author{Maurizio Giannotti}
\email{MGiannotti@barry.edu}
\affiliation{Physical Sciences, Barry University, 11300 NE 2nd Ave., Miami Shores, FL 33161, USA}

\author{Joerg Jaeckel}
\email{jjaeckel@thphys.uni-heidelberg.de}
\affiliation{Institut f\"ur theoretische Physik, Universit\"at Heidelberg,
Philosophenweg 16, 69120 Heidelberg, Germany}

\author{Alessandro Mirizzi}
\email{alessandro.mirizzi@ba.infn.it }
\affiliation{Dipartimento Interateneo di Fisica ``Michelangelo Merlin'', Via Amendola 173, 70126 Bari, Italy.}
\affiliation{Istituto Nazionale di Fisica Nucleare - Sezione di Bari,
Via Orabona 4, 70126 Bari, Italy.}


\begin{abstract}
The cumulative emission of Axion-Like Particles (ALPs) from all past core-collapse supernovae (SNe)
would lead to a diffuse flux with energies ${\mathcal O}(50)$ MeV. We use this to constrain ALPs featuring couplings to photons and to nucleons.
ALPs coupled only to photons are produced in the SN core via the Primakoff process,
and then converted into gamma rays in the Galactic magnetic field.
We set a bound on $g_{a\gamma} \lesssim 5 \times 10^{-10}$~GeV$^{-1}$ for $m_a \lesssim 10^{-11}$ eV, using recent measurements of the diffuse gamma-ray flux observed by the {\it Fermi}-LAT telescope.
However, if ALPs couple also with nucleons, their production rate in SN can be considerably enhanced due to the ALPs nucleon-nucleon bremsstrahlung process.
Assuming the largest ALP-nucleon coupling phenomenologically allowed, 
bounds on the diffuse gamma-ray flux lead to a much stronger  
$g_{a\gamma} \lesssim 6 \times 10^{-13}$~GeV$^{-1}$ for the same mass range.
If ALPs are heavier than $\sim$ keV, the decay into photons becomes significant, leading again to a 
diffuse gamma-ray flux.
In the case of only photon coupling, we find, e.g. 
$g_{a\gamma} \lesssim 5 \times 10^{-11}$~GeV$^{-1}$ for $m_a \sim 5$~keV. 
Allowing for a (maximal) coupling to nucleons, the limit improves to the level of 
$g_{a\gamma} \lesssim 10^{-19}$~GeV$^{-1}$ 
for  $m_a \sim 20$~MeV, which represents the strongest constraint to date. 
\end{abstract}

\maketitle

\section{Introduction}

Core-collapse supernovae (SNe) could be a cosmic factory of axions 
and axion-like particles (ALPs)~\cite{Raffelt:1996wa,Raffelt:2006cw,Fischer:2016cyd}, allowing one to probe these particles and their couplings with a variety of techniques.
In fact, the SN 1987A neutrino detection has been a milestone event also for axion physics~\cite{Turner:1987by,Burrows:1988ah,Burrows:1990pk,Raffelt:1987yt,Raffelt:1990yz,Keil:1996ju,Hanhart:2000ae,Chang:2018rso}. 
The dominant production channel for QCD axions is the axion nucleon-nucleon (NN) bremsstrahlung. 
Requiring that the axion emission does not 
reduce the SN neutrino burst significantly limits the nucleon-axion coupling to  
$g_{aN} \lesssim 10^{-9}$, the exact
bound depending on the specific axion model~\cite{Carenza:2019pxu}.
If ALPs couple only to photons, the dominant production mechanism is the Primakoff process. 
In this case, the energy loss argument leads to competitive limits compared with other observations only for masses $\gtrsim 50$~keV. However, also for small masses
the SN ALP flux can still lead to interesting constraints. In the case of light ALPs, the SN produced ALPs can 
convert into gamma rays in the magnetic field of the Milky Way~\cite{Grifols:1996id,Brockway:1996yr}. The lack of a gamma-ray signal in the
 Gamma-Ray Spectrometer (GRS) on the Solar Maximum Mission (SMM)
in coincidence with the observation of the neutrinos emitted from SN 1987A therefore provided a 
strong bound on ALPs coupling to photons~\cite{Grifols:1996id,Brockway:1996yr}. Notably for $m_a < 4 \times 10^{-10}$~eV the most recent analysis finds
 $g_{a\gamma} < 5.3 \times 10^{-12}$ GeV$^{-1}$~\cite{Payez:2014xsa}. 
 On the other hand, heavy ALPs ($m_a \sim {\mathcal O}(0.1-100)$~MeV) can decay into gamma rays on their route from the SN to Earth and the non-observation of a gamma-ray signal in coincidence with the SN 1987A implies the bound 
 $g_{a\gamma} < 3 \times 10^{-12}$ GeV$^{-1}$ at $m_{a}\sim 50$~MeV~\cite{Jaeckel:2017tud}.
 
The physics potential of a future SN explosion in improving these bounds has been also explored. In particular, it has been realized 
that if a Galactic SN would explode during the lifetime of the {\it Fermi}~satellite, one could improve significantly the 
previous constraints~\cite{Jaeckel:2017tud,Meyer:2016wrm}.
Furthermore, a search for
gamma-ray bursts from extragalactic SNe with
the Large Area Telescope aboard the {\it Fermi}~satellite (hereinafter {\it Fermi}-LAT)  has yielded the limit $g_{a\gamma} < 2.6 \times 10^{-11}$ GeV$^{-1}$, for ALP masses
$m_a < 3 \times 10^{-10}$~eV, under the assumption of at least one SN occurring in the detector field of view~\cite{Meyer:2020vzy}.
Also it has recently been discussed that if a SN explosion occurs within few hundred kiloparsecs from  Earth one may detect the 
emitted ALPs with an upgraded version of a next-generation helioscope~\cite{Ge:2020zww}.

However, (extra-)galactic SNe are rare and unpredictable events. Therefore, it is interesting to investigate if one can rely on a guaranteed SN ALP flux 
without waiting for the next Galactic explosion. Indeed, it has been shown in~\cite{Raffelt:2011ft} that the cumulative axion emission from all past core-collapse
SNe in the Universe would lead to a diffuse axion flux comparable with that of neutrinos. 
While the diffuse SN neutrino background is potentially detectable in Super-Kamiokande enriched with gadolinium~\cite{Beacom:2003nk,Horiuchi:2008jz}, the analysis of~\cite{Raffelt:2011ft} indicates that a detection of the diffuse SN axion flux is very challenging.
A possibility would be to exploit the same scheme applied to SN 1987A: axion production in the SN core and conversions into photons in the Milky Way magnetic field. 
However, crucially the analysis of Ref.~\cite{Raffelt:2011ft} focused on QCD axions with masses ${\mathcal O}$(meV).
Even though the axion production in SN is significant, because of the efficient axion NN bremsstrahlung process,  
the large axion mass hampers the conversions into photons in the Galactic magnetic field.
In contrast, in what follows we want to consider generic ALPs with mass and couplings completely unrelated.
In this case, there exist significant regions in the parameter space where we can have a large ALP production and sizable photon conversions. This then provides a gamma-ray signal which can be constrained by the diffuse gamma-ray background measured by {\it Fermi}-LAT. 

For heavier ALPs an alternative is to consider decays into photons~\cite{Jaeckel:2017tud,DeRocco:2019njg}. Indeed, in~\cite{DeRocco:2019njg} COMPTEL measurements~\cite{comptel3} have already been used to obtain limits from this signature of the Diffuse SN ALP Background (DSNALPB). Below we will return to this, allowing for larger nucleon couplings and also including newer measurements by {\it Fermi}-LAT. 

\bigskip

Let us briefly outline our plan for the next sections. In Sec.~\ref{sec:alpflux} we present the SN ALP flux for different ALP models
and calculate the diffuse SN ALP flux. In Sec.~\ref{sec:gammaray} we characterize the ALP-photon conversions in the Galactic magnetic field and  we present our bounds from the diffuse gamma-ray flux measured by {\it Fermi}-LAT.  In Sec.~\ref{sec:decay} we consider the constraints coming from the diffuse gamma-ray flux from the decay of heavy ALPs produced in SNe.
In Sec.~\ref{sec:future} we comment on the perspective for improvements in sensitivity through next generation gamma-ray experiments in the MeV energy range. Finally, Sec.~\ref{sec:conclusions} provides a summary of our results and conclusions.

\section{SN ALP fluxes}\label{sec:alpflux}

\subsection{ALP emission from SNe}

In the minimal scenario, ALPs have only a two-photon coupling, described by the 
 Lagrangian~\cite{Raffelt:1987im}
\begin{equation}
\label{mr}
{\cal L}_{a\gamma}=-\frac{1}{4} \, \gag
F_{\mu\nu}\tilde{F}^{\mu\nu}a=\gag \, {\bf E}\cdot{\bf B}\,a~.
\end{equation}
This interaction allows for ALP production  in a stellar medium primarily through the Primakoff process~\cite{Raffelt:1985nk}, in which thermal photons are converted into ALPs in the electrostatic field of ions, electrons and protons.
In order to calculate the ALP production rate (per volume) in a SN core via Primakoff process we closely follow~\cite{Payez:2014xsa}, finding
\begin{eqnarray}
\dfrac{d \dot n_a}{dE}&=&
\frac{g_{a\gamma}^{2}\xi^2\, T^3\,E^2}{8\pi^3\, \left( e^{E/T}-1\right) } \nonumber \\
& &\left[ \left( 1+\dfrac{\xi^2 T^2}{E^2}\right)  \ln(1+E^2/\xi^2T^2) -1 \right] \,.
\label{eq:axprod}
\end{eqnarray}
Here, $E$ is the photon energy, $T$ the temperature and $\xi^2={\kappa^2}/{4T^2}$ with $\kappa$ the inverse Debye screening length, describing 
the finite range of the electric field surrounding charged particles in the plasma.  
In order to get the total ALP production rate per unit energy one has 
to integrate Eq.~\eqref{eq:axprod} over the SN volume. 
As a reference, we consider an SN model with an 18 $M_{\odot}$ progenitor, simulated in spherical symmetry with the AGILE-BOLTZTRAN code~\cite{Mezzacappa:1993gn,Liebendoerfer:2002xn}.
We assume that the effect of the progenitor mass in the ALP flux is rather mild.
Indeed, in~\cite{Payez:2014xsa} some of us have  considered two different stellar models,
with progenitor mass of 
10 and 18 $M_{\odot}$. In these two cases the differences between different stellar models, e.g. in terms of peak temperatures and other nuclear matter properties relevant for the ALP production, are actually only of the order of a few percent.
 However, preliminary studies with heavier progenitor mass 
suggest the possibility of larger variations in ALP fluxes
(see, e.g.~\cite{Meyer:2020vzy}, for the case of 40 $M_{\odot}$ progenitor).

Assuming $m_a \ll T$, we find that the time integrated ALP spectrum is given, with excellent precision, by the analytical expression
\begin{equation}
\frac{dN^{x}_a}{dE} = C \left(\frac{g_{ax}}{g^{\rm ref}_{ax}}\right)^2
\left(\frac{E}{E_0}\right)^\beta \exp\left( -\frac{(\beta + 1) E}{E_0}\right) \,,
\label{eq:time-int-spec}
\end{equation}
where the values of the 
parameters $C$, $E_0$, and $\beta$ and the relevant reference couplings $g^{\rm ref}_{ax}$ for the different channels  $x$ are given in Table~\ref{tab:fitting}.
The spectrum described in Eq.~\eqref{eq:time-int-spec} is a typical quasi-thermal spectrum, with mean energy $E_0$ and index $\beta$ (in particular, $\beta=2$ would correspond to a perfectly thermal spectrum of ultrarelativistic particles).

 \begin{table}[!t]
\begin{center}
\begin{tabular}{lcccc}
\hline
 & 
$C$ [${\rm MeV}^{-1}$]
& $E_0$ [MeV] &$\beta$ &$g^{\rm ref}_{ax}$\\
\hline
\hline
$\gamma\to a$& $1.37 \times 10^{51}    $ & 122.3 &2.3 & $10^{-11}\,{\rm GeV}^{-1}$\\
$NN\to a$~~~&$9.08 \times 10^{55} $  & 103.2 & 2.2 & $10^{-9}$ \\
${\bar\nu}_e$ & $7.8  \times 10^{55} $ & 9.41 & 1.6 &~~~~~~~~~~N/A~~~~~~~~~~ \\
\hline
\end{tabular}
 \caption{Fitting parameters for the SN ALP spectrum from the Primakoff process and NN bremsstrahlung. 
 For comparison we also show the parameters corresponding to the  ${\bar \nu}_e$ spectrum.
 }
\label{tab:fitting}
\end{center}
\end{table}

If ALPs couple also with nucleons, the ALP NN bremsstrahlung process
\begin{eqnarray}
N_1 + N_2 \longrightarrow N_3 + N_4 + a\,,
\label{eq:brems}
\end{eqnarray}
provides another efficient production channel~\cite{Turner:1987by}.
In Eq.~\eqref{eq:brems}, $N_i $ are nucleons (protons or neutrons) and $ a $ is the ALP field.
The process \eqref{eq:brems} is induced by the ALP-nucleon interaction described by the following 
Lagrangian term~\cite{Carena:1988kr},
\begin{equation}\label{eq:axion_N_coupling}
\mathcal{L}_{a N}=\sum_{i=p,n} \frac{g_{a i}}{2m_N}\,\overline N_i\gamma_\mu\gamma_5 N_i  \partial^\mu  a,
\end{equation}
with $g_{a i}$ the ALP-nucleon couplings.
This process has been recently reevaluated in~\cite{Carenza:2019pxu}, including corrections beyond
the one-pion-exchange (OPE) approximation.  
In this case, assuming ALPs coupled only to protons, one finds the bound $g_{ap} \lesssim 1.2 \times 10^{-9}$,  required to avoid an excessive SN cooling that 
would have shortened  the duration observed SN 1987A neutrino burst.

Once again, the time-integrated spectrum is well represented by Eq.~(\ref{eq:time-int-spec}),
with fitting parameters given in Table~\ref{tab:fitting}. 
Indeed neutrino emission can also be described by such a spectral shape. The corresponding parameters are given in Table~\ref{tab:fitting} for comparison. Note that, this latter flux features a much lower average energy.
  
\begin{figure}[t!]
\vspace{0.cm}
\includegraphics[width=0.95\columnwidth]{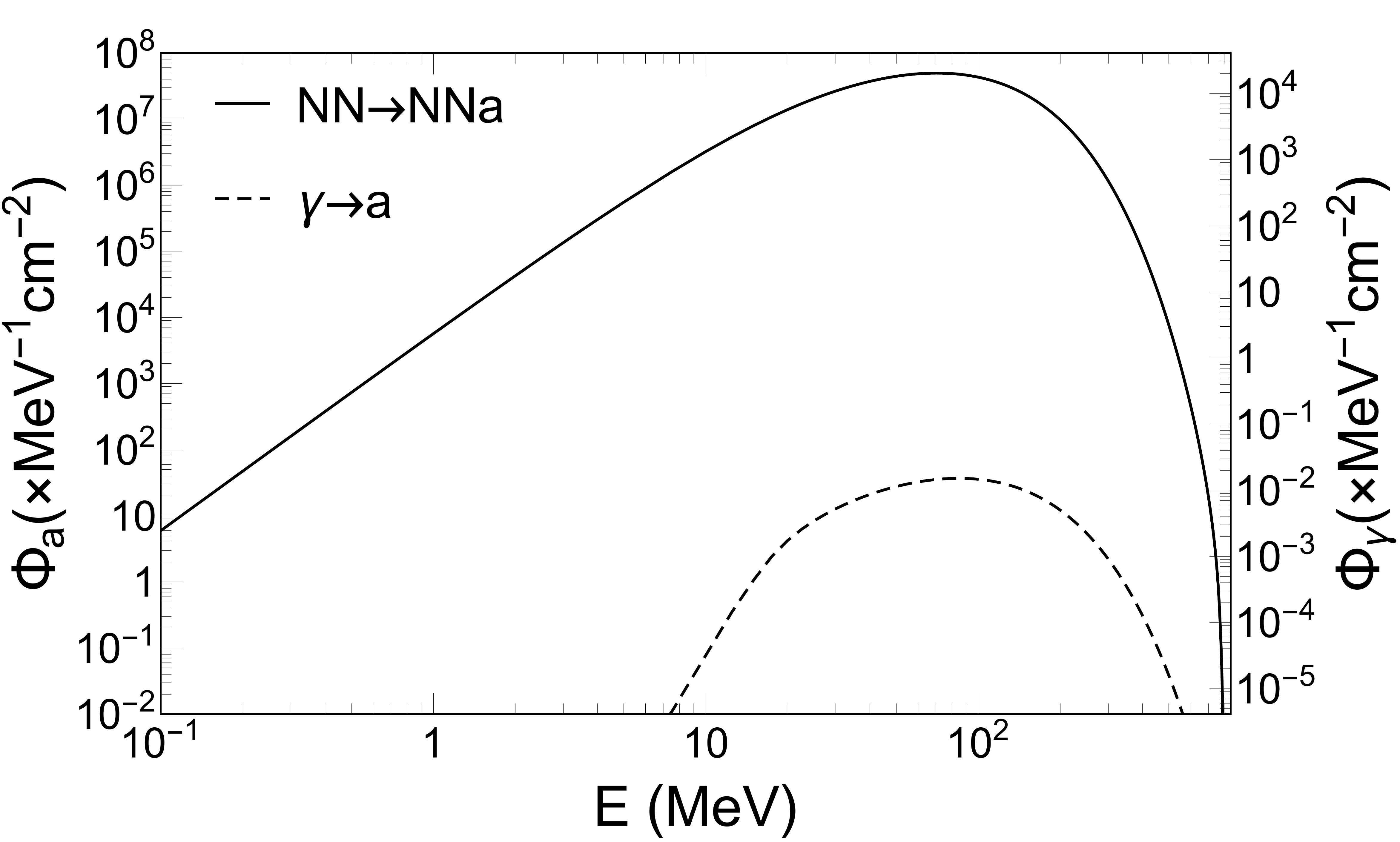}
\caption{SN ALP fluxes from the Primakoff process with $g_{a\gamma} = 5 \times 10^{-12}$ GeV$^{-1}$ (dashed curve) and NN bremsstrahlung
with  $g_{ap}= 10^{-9}$ (continuous curve). 
The label on the right vertical axis refers to the converted gamma-ray flux in the Galactic magnetic field assuming the direction of 
the SN 1987A (see the text for details). We assume $m_a \ll 10^{-11}$~eV.}
\label{fig:fluxes}
\end{figure}

In Fig.~\ref{fig:fluxes} we show an example of the SN ALP fluxes from the Primakoff process with $g_{a\gamma} = 5 \times 10^{-12}$ GeV$^{-1}$ (dashed curve) and NN bremsstrahlung
with $g_{ap}=  10^{-9}$ (continuous curve), i.e. we take both the couplings close to their respective individual limits.
We assumed $m_a \ll 10^{-11}$~eV in order to have seizable conversions into photons in the Galactic magnetic field (see below).
One realizes that both the fluxes are peaked at $E\sim 100$ MeV, but the flux from the NN process exceeds the one from the Primakoff process by 
7 orders of magnitude.
One thus expects that such huge fluxes would lead to an important gamma-ray flux, after ALPs conversion in Galactic magnetic field.

\bigskip 

In more general models, we could also allow for ALP interactions with electrons.
\footnote
{The ALP-photon coupling may also radiatively induce a coupling to electrons  (see, e.g.~\cite{Srednicki:1985xd,DiLuzio:2020wdo}).
This coupling is, however, strongly suppressed. 
Therefore, in our discussion of the ALP production, we ignore this effect and consider the ALP couplings with electrons and photons as completely unrelated. One may wonder if this is different when considering ALP decays. We will briefly discuss this at the end of Sect.~\ref{sec:decay} but the upshot is that the effect is always negligible compared to the decay into photons. 

}
An ALP-electron interaction leads to ALP production in the SN core via the Compton effect ($e^{-}+\gamma\rightarrow e^{-}+a$) and electron bremsstrahlung ($e^{-}+Ze\rightarrow e^{-}+a$). In the SN environment electrons are degenerate and the Compton effect is suppressed. Therefore, the relevant production channel is bremsstrahlung. This has the emissivity~\cite{Raffelt:1996wa}
\begin{equation}
\epsilon_{a}=1.26\times10^{40}{\rm erg}\, {\rm g}^{-1}{\rm s}^{-1}g_{ae}^{2}\left\langle\frac{Z^{2}}{A}\right\rangle\left(\frac{T}{30 \,\ {\rm MeV}}\right)^{4}F\;,
\end{equation}
where $F$ is a numerical factor of the order of unity. In order to respect other existing bounds on $g_{ae}$, we are forced to consider at most $g_{ae}\sim10^{-13}$
obtaining $\epsilon_{a}\sim10^{14} \, {\rm erg}\, {\rm g}^{-1}{\rm s}^{-1}$. Therefore this emissivity is orders of magnitude smaller than the $10^{19} \, {\rm erg}\, {\rm g}^{-1}{\rm s}^{-1}$, which is possible in the case of NN bremsstrahlung. Clearly this process can still be competitive if we consider a smaller ALP-nucleon coupling but we prefer to reduce the number of free parameters for the sake of clarity.

\subsection{Diffuse SN ALP background}

From the SN ALP flux described in the previous section, one can calculate the DSNALPB
from all past core-collapse SNe in the Universe, in analogy to the well-known prescription used in the neutrino case~\cite{Beacom:2010kk,Raffelt:2011ft}
\begin{equation}
\frac{d \phi_a (E_a)}{d E_a}\!\! =\!\! \int_0^{\infty} \! (1+z) \frac{dN_a(E_a(1+z))}{dE_a}[R_{SN}(z)] \bigg[ \bigg|c \frac{dt}{dz} \bigg| dz \bigg]
\label{eq:diffuse}
\end{equation}
where $z$ is the redshift. $R_{SN}(z)$
is the SN explosion rate, taken from~\cite{Priya:2017bmm}, with a total normalization for the core-collapse rate $R_{cc}=1.25 \times 10^{-4}$
yr$^{-1}$ Mpc$^{-3}$. 
Furthermore, $ |{dt}/{dz} |^{-1}= H_0(1+z)[\Omega_\Lambda+\Omega_M(1+z)^3]^{1/2}$ with the cosmological parameters 
$H_0= 67.4$ km s$^{-1}$ Mpc$^{-1}$, $\Omega_M=0.315$, $\Omega_\Lambda=0.685$~\cite{Aghanim:2018eyx}.
For simplicity we assume that the flux of ALPs from SNe of different masses is well represented by the ones we calculated in the SN model
used in the previous section. Given the mild dependence of the ALP flux on the SN progenitor mass, this assumption is consistent with the level of precision we are working at.

We can parametrize the obtained DSNALPB spectrum with the same spectral shape of Eq.~(\ref{eq:time-int-spec}). 
We show the fitting 
parameters in Table~\ref{table:dsnalpb}.

\begin{table}[!t]
\begin{center}
\begin{tabular}{lcccc}
\hline
 & 
$C$ [MeV$^{-1}\!\!$ cm$^{-2}$ s$^{-1}$]
& $E_0$ [MeV] &$\beta$ & ~~~~~$g^{\rm ref}_{ax}$~~~~~
\\
\hline
\hline
$\gamma\to a$ & $ 6.94 \times 10^{-6}  $ & 74.6 &1.75 & $10^{-11}\,{\rm GeV}^{-1}$
\\
$NN\to a$&$4.2 $  & 65.1 & 1.55 &$10^{-9}$ 
\\
\hline
\end{tabular}
 \caption{Fitting parameters of the DSNALPB spectrum from Primakoff process and NN bremsstrahlung.}
\label{table:dsnalpb}
\end{center}
\end{table}

\begin{figure}[t!]
\vspace{0.cm}
\includegraphics[width=0.8\columnwidth]{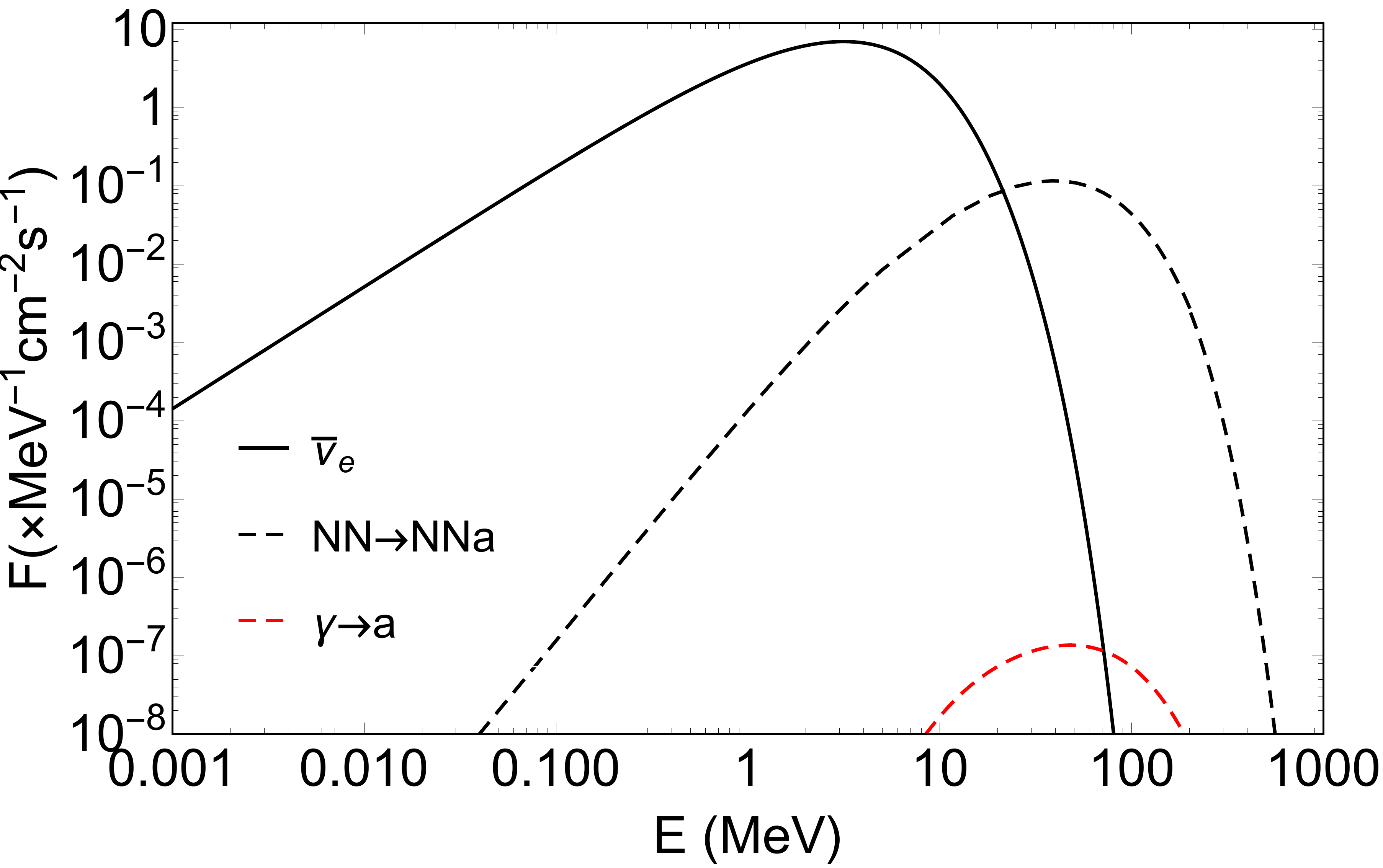}
\caption{DSNALPB fluxes from the Primakoff process with $g_{a\gamma} = 5.3 \times 10^{-12}$ GeV$^{-1}$ (red dashed curve) and NN bremsstrahlung
with  $g_{ap}=1.2 \times 10^{-9}$ (black dashed curve), compared with the diffuse ${\bar\nu}_e$ one (black continuous curve). 
 }
\label{fig:dsalpfl}
\end{figure}

In Fig.~\ref{fig:dsalpfl} we show the DSNALPB fluxes for a pure photon coupling $g_{a\gamma} = 5 \times 10^{-12}$ GeV$^{-1}$ (red dashed curve) and a pure nucleon coupling $g_{ap}= 10^{-9}$ (black dashed curve), compared with the ${\bar\nu}_e$ one (continuous curve).
We can see that the ALP flux is peaked at  higher energies ($E\sim 50$ MeV) with respect to the neutrino one. 
Moreover, with the used ALP-nucleon coupling, the integrated total flux is actually larger than that of neutrinos (note that we have to integrate over energy which is shown logarithmically in Fig.~\ref{fig:dsalpfl}, slightly obscuring the fact that the relevant range is bigger for ALPs).

\section{Gamma-ray flux from DSNALPB conversions}\label{sec:gammaray}

\subsection{ALP--photon conversions in the Milky Way}
Once ALPs are produced in a SN core, they can easily escape the star since their mean free path in stellar matter is sufficiently large for the values of the coupling $g_{a\gamma}$ that we are considering~\cite{Brockway:1996yr}.
Then, they will propagate until they reach the Milky Way. There they can convert into photons in the Galactic magnetic field. 
Indeed, the Lagrangian given in Eq.~\eqref{mr} would trigger ALP--photon oscillations in external magnetic fields.

The problem of photon-ALP conversions  simplifies  if one restricts the attention to the case in which ${\bf B}$ is homogeneous. We denote by ${\bf B}_T$ the transverse magnetic field, namely its component in the plane normal to the
photon beam direction. The linear photon polarization state parallel to the transverse field direction ${\bf B}_T$ is then denoted by $A_{\parallel}$ and the orthogonal one by $A_{\perp}$. 
The  component $A_{\perp}$ decouples,
while the probability for a photon emitted in the state $A_{\parallel}$ to oscillate into an ALP after traveling a distance $d$ is given by~\cite{Raffelt:1987im},
\begin{equation}
P_{\gamma \to a} 
= (\Delta_{a \gamma} d)^2 \frac{\sin^2(\Delta_{\rm osc} d/2)}{(\Delta_{\rm osc} d/2)^2} \,\ .
\label{conv}
\end{equation}
Here, the oscillation wave number is ~\cite{Raffelt:1987im}
\begin{equation}
\Delta_{\rm osc} \equiv \left[(\Delta_{a} - \Delta_{\rm pl})^2 + 4 \Delta_{a \gamma}^2 \right]^{1/2} \,\ ,
\end{equation}
with  $\Delta_{a\gamma} \equiv {g_{a\gamma} B_T}/{2} $ and $\Delta_a \equiv - {m_a^2}/{2E}$. The term $\Delta_{\rm pl} \equiv -{\omega^2_{\rm pl}}/{2E}$ accounts for plasma effects, where $\omega_{\rm pl}$ is the  plasma frequency (see, e.g.,~\cite{Kopf:1997mv})
expressed as a function of the free electron density in the medium $n_e$ as $\omega_{\rm pl} 
= (4 \pi \alpha n_e/m_e)^{1/2}\simeq 3.69 \times 10^{- 11} \sqrt{n_e /{\rm cm}^{- 3}} \, {\rm eV}$. For our benchmark values of the relevant parameters, numerically we find 
\begin{eqnarray}  
\Delta_{a\gamma}\!\!\! &\simeq &\!\!\! 1.5\times10^{-2} \left(\frac{g_{a\gamma}}{10^{-11}\textrm{GeV}^{-1}} \right)
\left(\frac{B_T}{10^{-6}\,\rm G}\right) {\rm kpc}^{-1}
\nonumber\,,\\  
\Delta_a\!\!\! &\simeq &\!\!\!
 -7.8 \times 10^{-4} \left(\frac{m_a}{10^{-11} 
        {\rm eV}}\right)^2 \left(\frac{E}{10 \,\ {\rm MeV}} \right)^{-1} {\rm kpc}^{-1}
\nonumber\,,\\  
\Delta_{\rm pl}\!\!\! &\simeq &\!\!\!
  -1.1\times10^{-5}\left(\frac{E}{10 \,\ {\rm MeV}}\right)^{-1}
         \left(\frac{n_e}{10^{-3} \,{\rm cm}^{-3}}\right) {\rm kpc}^{-1}
\nonumber\, .
\label{eq:Delta0}\end{eqnarray}
One realizes that for $m_{a}\ll 10^{-11}\,{\rm eV}$ and $E\gtrsim 10$~MeV, this becomes energy-independent, $P_{\gamma \to a} \simeq (\Delta_{a \gamma} d)^2$, since $\Delta_{a\gamma} \gg \Delta_a, \Delta_{\rm pl}$.

\begin{figure}[t!]
\vspace{0.cm}
\includegraphics[width=0.95\columnwidth]{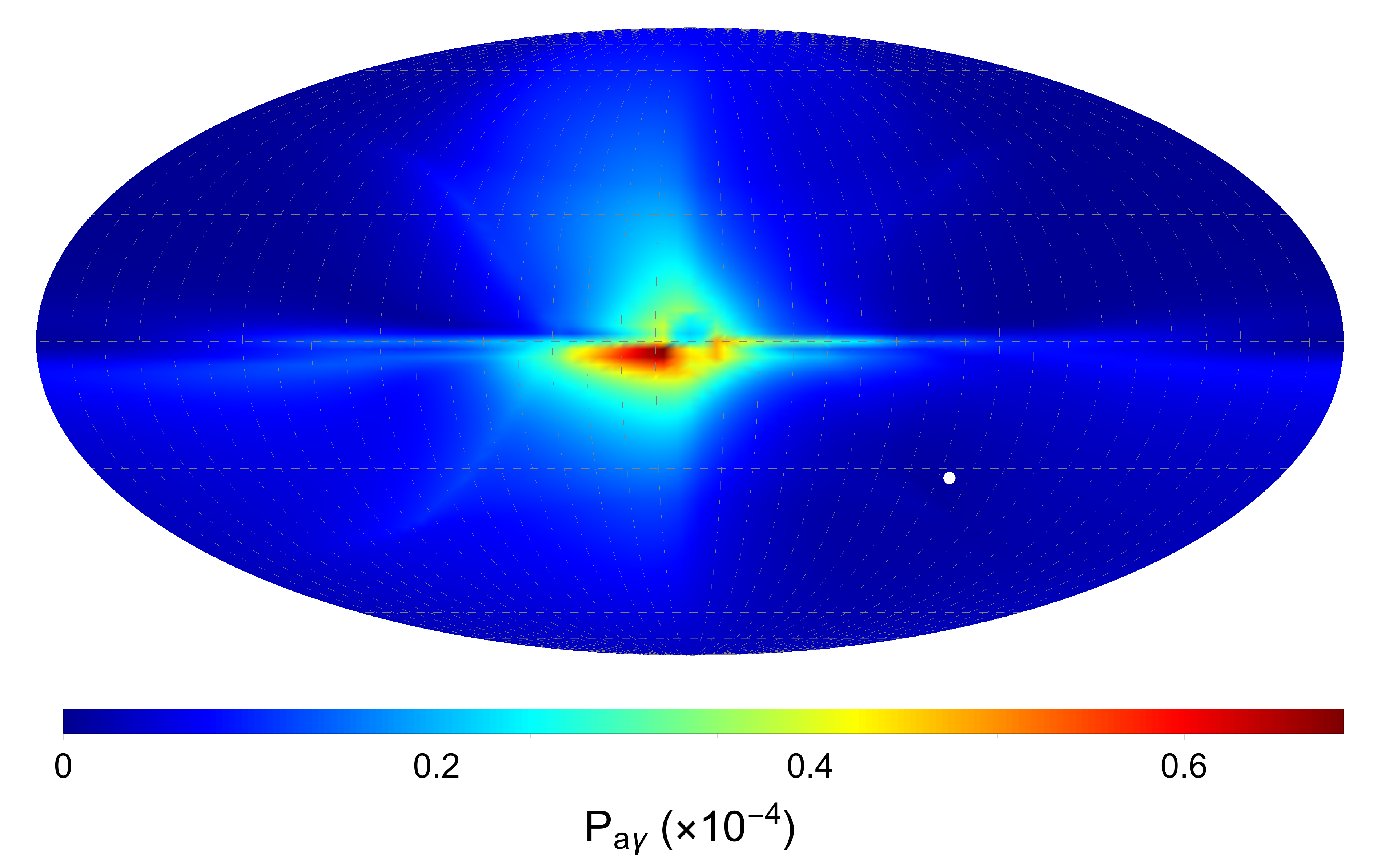}
\caption{Sky map in Galactic coordinates of the $a\to \gamma$ conversion probability, starting from a pure ALPs beam at the outside boundary of the Galaxy, for the Jansson and Farrar magnetic field model derived in~\cite{Jansson:2012pc}. We have taken the energy to be $E = 50$ MeV, the coupling $g_{a\gamma} = 3 \times  10^{-13}$GeV$^{-1}$ and $m_a \ll 10^{-11}$ eV. The white circle represents the Galactic coordinates  of the SN 1987A.
}
\label{fig:gal}
\end{figure}

Measurements of the Faraday rotation based on pulsar observations have shown that the regular component of  the Galactic
magnetic field  is parallel to the Galactic plane, with a typical strength $B \simeq$ a few~$\mu{\rm G}$,
and radial coherence length  $l_r \simeq10~{\rm kpc}$~\cite{Beck:2008ty}. Inside the Milky Way disk the electron density is \mbox{$n_e \simeq 1.1 \times 10^{-2}~{\rm cm}^{-3}$}, resulting in a plasma frequency ${\omega}_{\rm pl} \simeq 4.1 \times 10^{-12}~{\rm eV}$.
Among the possible magnetic field models proposed in the literature, 
we take the  Jansson and Farrar model~\cite{Jansson:2012pc} as our benchmark, with the updated parameters  given in Table~C.2 of~\cite{Adam:2016bgn} (``Jansson12c'' ordered fields).
Because of the presence of a rather structured behavior in the Galactic magnetic field,
the propagation of ALPs in the Galaxy is clearly a truly 3-dimensional problem. Because of the variations of the direction of ${\bf B}$ the same photon
polarization states play the role of either $A_{\parallel}$ and  $A_{\perp}$ in
different domains. 
We have closely followed the technique described in Ref.~\cite{Horns:2012kw} (to which we direct the reader for more details) to
solve the beam propagation equation along a Galactic line of sight.
Finally, the differential photon flux per unit energy arriving at Earth is given by,
\begin{equation}
\frac{d\Phi_{\gamma}}{dE}= \frac{1}{4 \pi d^2} \frac{d\dot{N}_a}{dE} \times P_{a\gamma} \,,\label{eq:diffphotonflux}
\end{equation}
where $d$ is the SN distance.

An illustrative sky map of the line-of-sight dependent probability for an ALP starting at the edge of the Galaxy to convert into a photon at Earth is shown in 
Fig.~\ref{fig:gal} for our chosen reference Jansson and Farrar magnetic field model. 
The probability of $a \to \gamma$ conversion is generally larger toward the Galactic center due to the presence of the X-shaped field and to the large vertical scale height of the halo field.
\footnote{ 
It is worth mentioning 
that the choice of magnetic field model 
leads to some uncertainty in the 
ALP flux. For example, the Pshirkov \emph{et al.} model~\cite{Pshirkov:2011um} used in~\cite{Payez:2014xsa}
would predict a magnetic field larger by more than a factor 3 for $d \lesssim 6$ kpc, and thus a larger 
ALP flux. Therefore, our choice is conservative in this respect.}
For light ALPs, when the mass effects in the oscillation  probability can be ignored, the photon spectrum has the shape
\begin{align}
\label{eq:shape1}
\frac{dF_{\gamma}}{dE}=\left( a(E) g_{a\gamma,11}^2+b(E) g_{aN,9}^2 \right) g_{a\gamma,11}^{2},
\end{align}
with $g_{a\gamma,11}=g_{a\gamma}\cdot 10^{11}\,{\rm GeV}$, $g_{aN,9}=g_{aN} \cdot 10^{9}$. Both 
functions $a$ and $b$ have the shape given in Eq.~\eqref{eq:time-int-spec}.
Specifically,
\begin{align}
\label{eq:shape2}
a(E)=A\left(\frac{E}{E_0}\right)^\beta \exp\left( -\frac{(\beta + 1) E}{E_0}\right)  \, ,
\end{align}
with $A=2.7\times 10^{-9}$ $\textrm{MeV}^{-1}\,{\rm cm^{-2}\,s^{-1}} $, and 
\begin{align}
\label{eq:}
b(E)=B\left(\frac{E}{E_0}\right)^\beta \exp\left( -\frac{(\beta + 1) E}{E_0}\right) \, ,
\end{align}
with $B=7.1\times 10^{-4}$
$\textrm{MeV}^{-1}\,{\rm cm^{-2}\,s^{-1}} $.
All the other parameters are as given in Table~\ref{table:dsnalpb}.

\subsection{Limits from SN 1987A}

Let us briefly comment on the limit obtainable from SN 1987A.
Assuming a SN with the position coincident with SN 1987A (corresponding to a distance $d=50$ kpc with  a Galactic latitude
$b=-32.1^\circ$ and longitude $l=279.5^\circ$)
 one finds that the energy-independent ALP-photon conversion probability would scale as
$P_{a\gamma}=0.15 \times g_{10}^2$. 
The resultant photon flux is quoted in the right $y$-label in Fig.~\ref{fig:fluxes}. 
Following~\cite{Payez:2014xsa} one should impose that in the energy window $E\sim  [25,100]$ MeV, the time-integrated
photon flux (over 10 s) should be smaller than $0.6$ cm$^{-2}$, in order not to exceed the background measured by the
the Gamma-Ray Spectrometer (GRS) on the Solar Maximum Mission. 
Fixing $g_{ap}$ at the previous value we find an upper bound on 
$g_{a\gamma} <3.4 \times 10^{-15}$~GeV$^{-1}$. Therefore, assuming a value of $g_{ap}$ close to the bound allowed by SN 1987A energy loss, improves the limit 
on $g_{a\gamma}$ by three orders of magnitudes compared to the case of ALPs coupled only to photons.

\subsection{Limits from DSNALPB}

As already suggested in~\cite{Raffelt:2011ft} one can also place a bound on the ALP coupling, comparing the diffuse gamma-ray flux coming from
ALP conversions with a measurement of the total diffuse gamma-ray cosmic background. 

The diffuse gamma-ray background associated 
with the DSNALPB can be obtained by convolving the ALP flux with the conversion probability
in the Galactic magnetic field. In principle this yields a map with a strong dependence on the direction in the sky as shown in Fig.~\ref{fig:gal}. 
In the following we make use only of the average flux. A detailed statistical analysis comparing the angular pattern with the observational data is left to future work.

The most recent measurement of the diffuse gamma-ray background at ${\mathcal O}(100)$ MeV has been performed by {\it Fermi}-LAT~\cite{Ackermann:2014usa}. The so-called isotropic diffuse gamma-ray background, or IGRB, is obtained by fitting the gamma-ray sky with an isotropic spatial template on top of known Galactic and extragalactic gamma-ray emission components, both diffuse and point-like. 
The IGRB is thus supposed to be the result of the superposition of Galactic and extragalactic contributions to the cosmic diffuse emission, not accounted for by known gamma-ray emission processes. These contributions may come from faint-luminosity populations of astrophysical objects such as active galactic nuclei (AGN) and external galaxies\footnote{Note that, the IGRB measurement (intensity and energy spectrum), therefore, strongly depends on the ability of the instrument to resolve individual point-like sources.} but also from some more exotic sources such as dark matter decay and annihilation or in our case ALPs.  
To be conservative, in the following, we take the whole of the IGRB as the maximal contribution allowed for ALPs to set our limit on the coupling.

For the present analysis we use the latest IGRB measurement provided by the {\it Fermi}-LAT Collaboration,\footnote{File \texttt{iso\_P8R3\_ULTRACLEAN\_V3\_v1.txt}, available at \url{https://fermi.gsfc.nasa.gov/ssc/data/access/lat/BackgroundModels.html}}  making use of Pass 8 R3 processed data (8-yr dataset), for the ULTRACLEANVETO event class selection (which minimizes the contamination from cosmic-ray events).
For $ E \gtrsim 50$ MeV, one can fit the IGRB measured flux with a power law
\begin{equation}
\Phi_{\gamma}(E) = 2.2 \times 10^{-3} E^{-2.2} \,\ \textrm{MeV}^{-1} \,\ \textrm{cm}^{-1} \,\ \textrm{s}^{-1} \,\ \textrm{sr}^{-1} \,\ .
\label{eq:fermi}
\end{equation}
Both the deviations from this fit as well as the experimental (statistical) errors are reasonably small in the range relevant to us. Since we use the whole flux to set our limits, these uncertainties can be neglected and we are justified in using this fit.  We stress, however, that systematic uncertainties due to the modeling of the Galactic diffuse emission may be important, and as shown in~\cite{Ackermann:2014usa} they amount to about +15\% on the integrated IGRB flux. Properly including systematics of the Galactic diffuse emission is beyond the scope of the current paper, and will be deferred to a future work, where a more complete analysis of {\it Fermi}-LAT data is planned to be performed.

\begin{figure}[t!]
\vspace{0.cm}
\includegraphics[width=0.95\columnwidth]{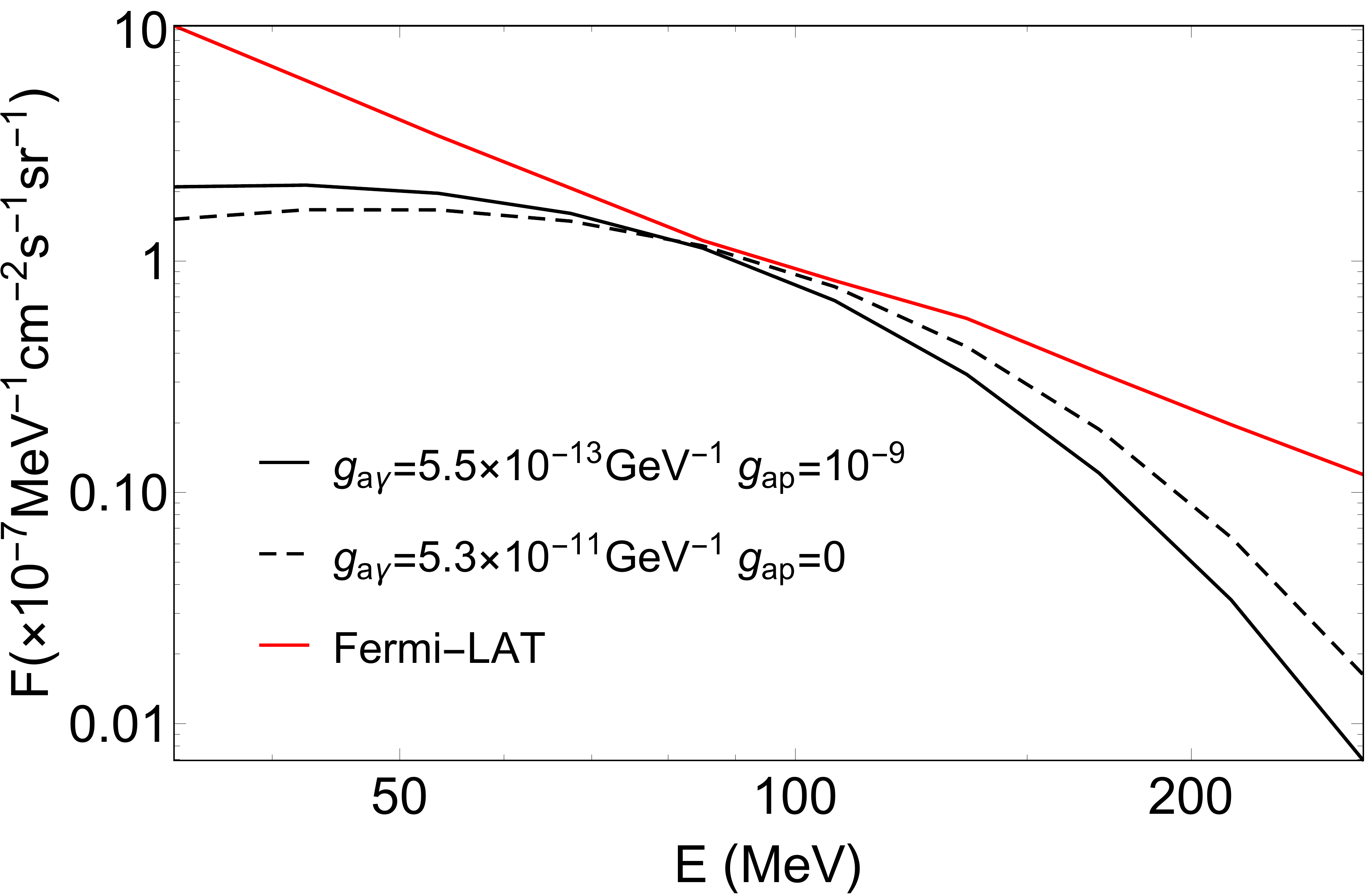}
\caption{Comparison of the diffuse gamma-ray flux from SN ALP conversions with the  diffuse gamma-ray 
flux measured by the {\it Fermi}-LAT experiment.}
\label{fig:fermi}
\end{figure}

In Fig.~\ref{fig:fermi} we compare the diffuse gamma-ray flux measured by the {\it Fermi}-LAT experiment with the gamma-ray flux
expected from conversions of DSNALPB. We consider $m_a \ll 10^{-11}$ eV.  
We see that for $g_{ap}=0$ one gets a bound 
$g_{a\gamma} \lesssim 5.3 \times 10^{-11}$ GeV$^{-1}$.
If we consider also a non vanishing ALP-proton coupling, assuming it at most 
$g_{ap}=10^{-9}$, we get  $g_{a\gamma} \lesssim 5.5 \times 10^{-13}$GeV$^{-1}$.  

\begin{figure}[t!]
\vspace{0.cm}
\includegraphics[width=0.99\columnwidth]{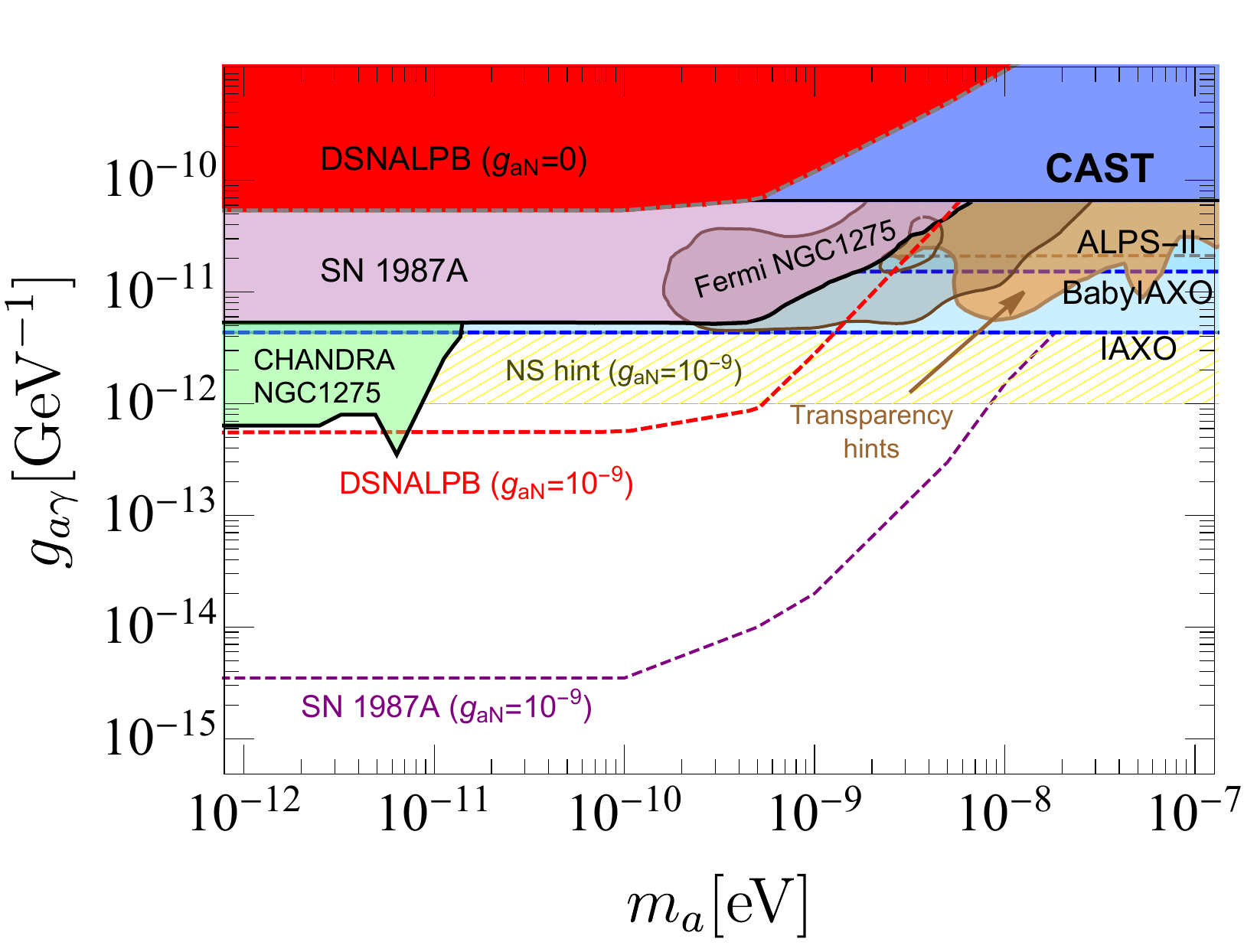}
\caption{
Summary plot of constraints for light ALPs. Our new limits from the DSNALPB are shown as the red area and the red dashed line for the pure photon coupling and the maximal nucleon coupling, respectively. The purple area gives the pure photon limit from SN 1987A~\cite{Payez:2014xsa} and the dashed line indicates the improvements possible with a maximal nucleon coupling. We also show the limits from CAST~\cite{Anastassopoulos:2017ftl} and the sensitivities of Baby-IAXO~\cite{Abeln:2020ywv}, IAXO~\cite{Armengaud:2019uso} and ALPS II~\cite{Bahre:2013ywa}. Observations made by CHANDRA~\cite{Reynolds:2019uqt} and {\it Fermi-}LAT~\cite{TheFermi-LAT:2016zue} of NGC 1275 constrain the  green and darker purple regions. The brown area delimits the hint of the anomalous transparency of the Universe for gamma rays~\cite{Meyer:2013pny,Kohri:2017ljt}. The hint~\cite{Buschmann:2019pfp} from X-ray observations of neutron stars is shown in yellow. 
} 
\label{fig:overview}
\end{figure}

In Fig.~\ref{fig:overview} we compare the bound on light ALPs from the DSNALPB with other constraints on the ALP parameter space.
We note that the DSNALPB bound is comparable or better than the CAST bound from solar ALPs~\cite{Anastassopoulos:2017ftl}, depending on the presence of the
ALP-nucleon coupling.
However, when compared with the SN 1987A bound the DSNALPB one is always less stringent. This is not surprising since 
one expects a much larger ALP flux from the SN 1987A than from the cumulative SN explosions. 
Nevertheless, due to the criticisms sometimes expressed towards the SN 1987A ALP bound, it is important to have  independent constraints
based on a fresh set of data.

\subsection{A hint from neutron stars}
It was recently pointed out in~\cite{Buschmann:2019pfp} that X-ray emissions of some isolated NS may be hinting at an ALP coupled to photons and to neutrons (see also \cite{Fortin:2018ehg}).

To compare with our bounds we first note that their result depends on the coupling to neutrons. Our limits are computed using a coupling to protons but up to a numerical factor close to $1$ they also apply to a coupling to neutrons.
In Fig.~\ref{fig:overview}, we show the suggested region in our plot under the assumption that the coupling to nucleons is maximal. Thus, the hinted region should be compared with the curves from DSNALPB and SN 1987A that include the corresponding couplings to nucleons. 
As we can see, a sizable part of the region hinted at by the NS observations is already excluded by the spectrum of the diffuse cosmic background as measured by {\it Fermi}-LAT and even a larger part by the non-observation of a gamma-ray flux by the SMM during SN 1987A.

\section{Gamma-ray flux from DSNALPB decay}\label{sec:decay}

Fundamentally, ALPs are unstable.
They can decay into two photons and, if massive enough, into other fields they are coupled to. 
The decay rate into two photons~\cite{Raffelt:2006rj,Carenza:2019vzg} 
\begin{equation}
\Gamma_{a\gamma\gamma}= \frac{g_{a\gamma}^2 m_a^3}{64 \pi} \,\ ,
\label{eq:decayr}
\end{equation}
is extremely small for light ALPs. 
For example, according to Eq.~\eqref{eq:decayr}, an ALP of mass $1\,$eV and coupling to photons $g_{a\gamma}=10^{-10}\,{\rm GeV^{-1}}$ has a lifetime 
of $1.3\times10^{25}\,$s, several orders of magnitude longer than the age of the Universe.
That justifies neglecting the ALP decay in our previous discussion.

Yet, the decay rate grows rapidly with the mass. 
Indeed, ALPs with a mass above a few keV have a non-negligible probability of decaying into photons on their way to Earth, thereby contributing to the gamma-ray background. 
In this section, we discuss this signature, which has also already been employed in~\cite{DeRocco:2019njg}.

Concretely, Eq.~\eqref{eq:decayr} results in a decay length for the ALPs given by
\begin{eqnarray}
l_{\rm ALP} &=& \frac{\gamma v}{\Gamma_{a\gamma\gamma}} = 
\frac{E_a}{m_a}\sqrt{1-\frac{m_a^2}{E_a^2}}  \frac{64 \pi}{g_{a\gamma}^2 m_a^3} \nonumber \\
&\simeq & 4 \times 10^{-3} \,\ \textrm{ly} \left(\frac{E_a}{100 \,\ \textrm{MeV}} \right) \left(\frac{10 \,\ \textrm{MeV}}{m_a} \right)^4 \nonumber \\
& & \qquad\qquad\qquad\times\left(\frac{10^{-10} \,\ \textrm{GeV}^{-1}}{g_{a\gamma}} \right)^2 \,\ .
\end{eqnarray}
The photon flux from decaying ALPs from a SN   can be expressed as
\begin{equation}
\frac{d N_\gamma(E_\gamma)}{d E_\gamma} =2 \times [1-\exp(-d/l_{\rm ALP})] \frac{dN_a(E_a)}{dE_a} \, ,
\label{eq:decay}
\end{equation}
where we use the approximation $E_a= 2 E_\gamma$. 

Note that, in the range of couplings and masses we are considering here, the decay length is sufficiently large for us to assume that all ALPs decay in a region transparent to photons, $l_{\rm ALP} \gg 3\times10^{12} {\rm cm}$ (see~\cite{Kazanas:2014mca}).

If $l_{\rm ALP} \lesssim 1/H_0$ a sizable fraction of ALPs from cosmological distances decay before they reach Earth.
In this case we expect a diffuse gamma-ray flux originating from the decay of heavy ALPs into photons.
This can be obtained by inserting the photon flux, Eq.~(\ref{eq:decay}), instead of 
the ALP flux into the cumulative
flux formula, Eq.~(\ref{eq:diffuse}). 

\begin{figure}[t!]
\vspace{0.cm}
\includegraphics[width=0.95\columnwidth]{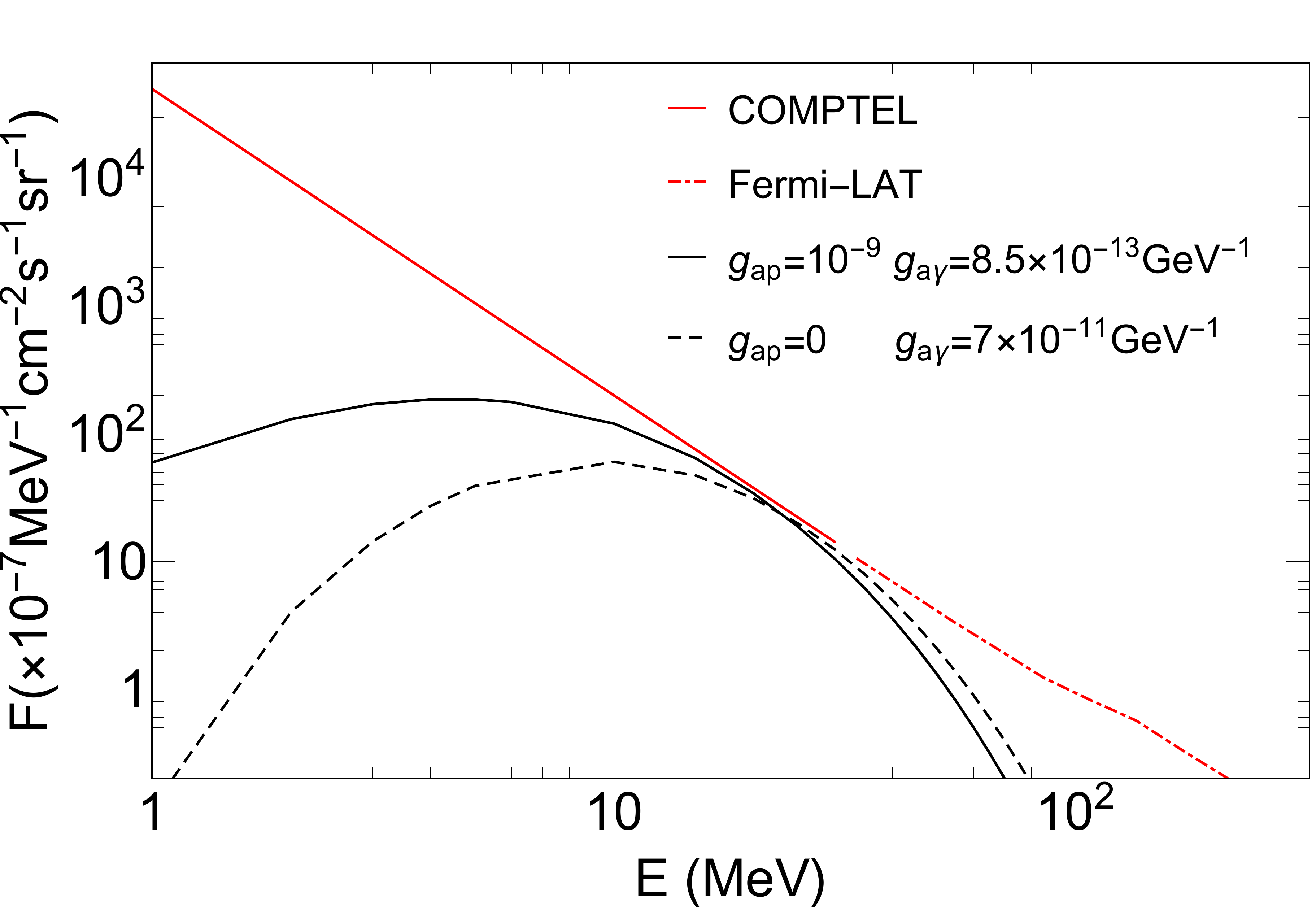}
\caption{Comparison of the diffuse gamma-ray flux from heavy SN ALP decay  with the  diffuse gamma-ray 
flux measured by the  {\it Fermi}-LAT and COMPTEL experiments. We consider an ALP with a mass $m_a=5$~keV.
}
\label{fig:comptel}
\end{figure}

In order to get a bound on the coupling $g_{a\gamma}$ we compare the diffuse photon flux generated by ALP decays
with the diffuse flux measured by {\it Fermi}-LAT at energies $E \gtrsim 30$~MeV, as given in Eq.~(\ref{eq:fermi}).
At lower energies ($0.8-30$ MeV) we consider the  experimental gamma-ray background flux measured by the COMPTEL experiment~\cite{comptel,comptel2}
\begin{equation}
\phi_\gamma(E)=1.05 \times 10^{-4}\left(\frac{E}{5 \,\ \textrm{MeV}}\right)^{-2.4}  \,\ \textrm{cm}^{-1} \,\ \textrm{s}^{-1} \,\ \textrm{sr}^{-1} \,\ .
\end{equation}
The flux limits from {\it Fermi}-LAT and COMPTEL together with two exemplary fluxes for an ALP with a mass $m_a=5$~keV are shown in Fig.~\ref{fig:comptel}.

\begin{figure}[t!]
\vspace{0.cm}
\includegraphics[width=0.99\columnwidth]{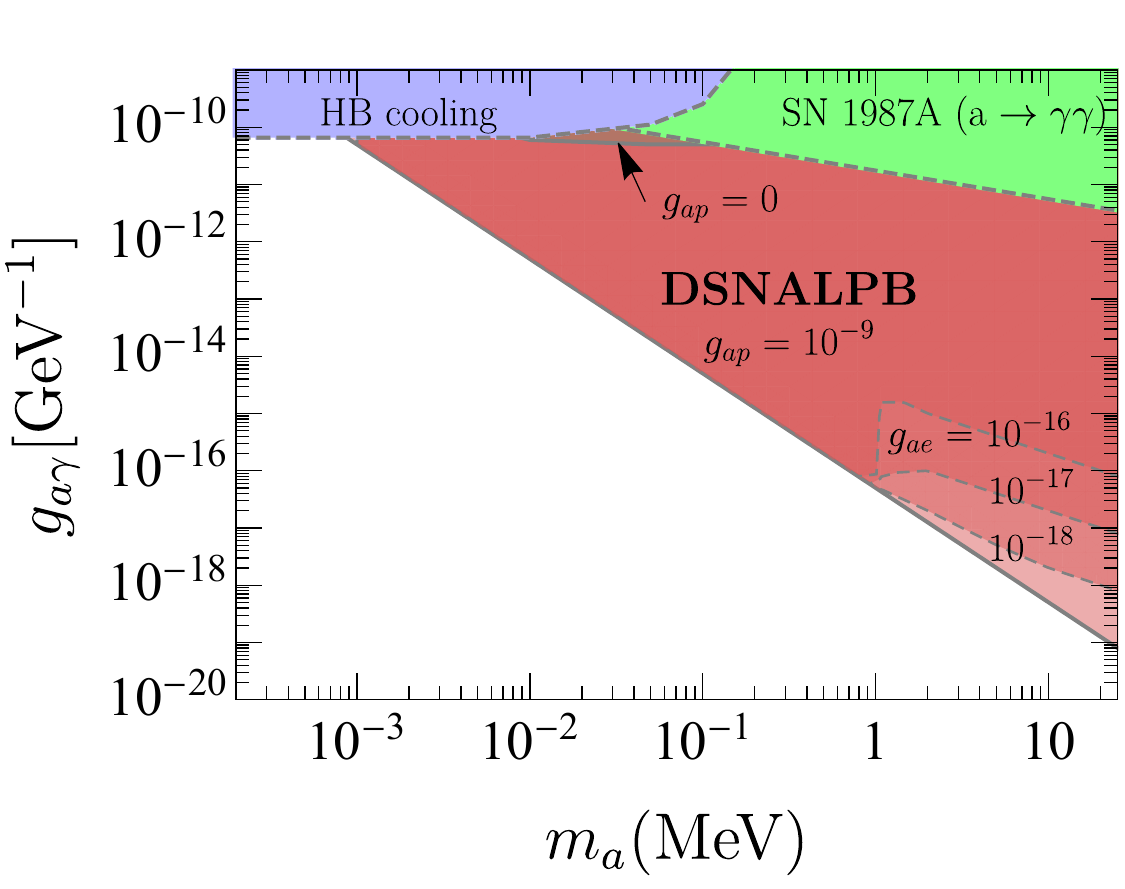}
\caption{Bounds on $g_{a\gamma}$ for decaying  heavy ALPs from the diffuse  SN gamma-ray flux. 
We show the case with $g_{ap}=0$ and $g_{ap}=10^{-9}$. In the latter case we also indicate the results for different values of the electron coupling $g_{ae}$.
For comparison we show the limits on the pure photon coupling from horizontal branch 
stars~\cite{Carenza:2020zil} and from SN 1987A~\cite{Jaeckel:2017tud}.
}
\label{fig:bound_massive}
\end{figure}

In Fig.~\ref{fig:bound_massive} we show the resulting bound on $g_{a\gamma}$  assuming massive ALPs coupling only with photons or 
with photons, nucleons  
and electrons.

For the pure photon coupling the resulting limit improves on existing limits in a sizable range from roughly $10$~keV to about $100$~keV. For example, at $m_a \gtrsim 5$~keV the resulting limit is roughly $g_{a\gamma} \lesssim 5 \times 10^{-11}$~GeV$^{-1}$.
We notice that our bound is comparable to but slightly better than
the one from the diffuse gamma-ray flux of decaying ALPs emitted by SNe presented in~\cite{DeRocco:2019njg}, most likely due to our use of newer data from {\it Fermi}-LAT.
This also has to be compared with the 
the bound placed by horizontal branch stars in globular clusters, i.e.  
$g_{a\gamma} \lesssim 6 \times 10^{-11}$~GeV$^{-1}$ for $m_a \lesssim 10$ keV~\cite{Carenza:2020zil}. 
At larger masses the bound on photon coupling is dominated by the one obtained from decaying ALPs from SN 1987A~\cite{Jaeckel:2017tud}.
The advantage of the DSNALPB over the limit from SN 1987A is due to the cosmological baseline $\sim 1/H_0$ involved in the diffuse flux calculation. One can therefore have decays also for smaller ALP masses, which have a longer lifetime, see Eq.~\eqref{eq:decay}.

If we allow for a nucleon coupling as large as $g_{ap}=10^{-9}$ we see that the corresponding bound on $g_{a\gamma}$ is significantly
improved, e.g. for  $m_a \gtrsim 5$~keV it reaches $g_{a\gamma} \lesssim 6 \times 10^{-13}$~GeV$^{-1}$ and continues to improve to a level $\sim 10^{-19}\,{\rm GeV}^{-1}$ at $\sim 20\,{\rm MeV}$, stronger than all other existing constraints. We also mention that ALPs  with masses in the keV-MeV range would have a profound impact on the cosmological evolution of
our Universe, in particular on the abundance of light elements produced during Big Bang Nucleosynthesis.
The resulting limits are complementary to the astrophysical ones we are discussing.
However, a potential drawback is that altering the cosmological history may potentially weaken or even fully invalidate
these bounds (see \cite{Depta:2020wmr} for a recent study).

ALPs with a mass $m_{a}>1\,{\rm MeV}$ can also decay into $e^{+}e^{-}$ pairs. This effect tends to reduce the decay length as well as the total number of photons produced from ALP decays. 
Considering only the electron coupling the length for the decay into electrons is given by~\cite{Altmann:1995bw},
\begin{equation}
\begin{split}
l_{e}=&\frac{\gamma v}{\Gamma_{a\rightarrow e^{+}e^{-}}}=1.39\times10^{13}{\rm \,cm}\left(\frac{10^{-13}}{g_{ae}}\right)^{2}\\
&\left(\frac{50 {\,\rm MeV}}{m_{a}}\right)^{2}\left(\frac{E}{70{\,\rm MeV}}\right)
\left[\frac{1-\left(\frac{m_{a}}{E}\right)^{2}}{1-\left(\frac{1{\,\rm MeV}}{m_{a}}\right)^{2}}\right]^{1/2}\;.
\label{eq:decayel}
\end{split}
\end{equation}
The term inside the square brackets is of order unity unless the ALP mass is close to the threshold allowing for a decay into electrons and positrons, or the ALP is non-relativistic.

Even when the decay channel into electrons is open, it is easy to verify that the axion decay length is always larger than $3\times10^{12} {\rm cm}$ (cf.~\cite{Kazanas:2014mca}). For example, according to Eq.~\eqref{eq:decayel}, an ALP of mass $50\,$MeV, energy of $70\,$MeV and a coupling to electrons $g_{ae}=10^{-13}$ has a decay length of $\sim 4\times10^{13}\,$cm. Therefore, in the following we will assume that all ALPs decay outside of the SN envelope.

Then, in order to count the number of photons, we include the branching ratio for the decay into photons as $\Gamma_{\gamma}/(\Gamma_{\gamma}+\Gamma_{e})$.
The result of this calculation is shown in Fig.~\ref{fig:bound_massive} as the dashed curves with the labels indicating the different values of $g_{ae}$.

{As the effect of an electron coupling is quite significant, let us briefly comment on the impact of an electron coupling that is loop induced from the photon coupling.
Following~\cite{Srednicki:1985xd,DiLuzio:2020wdo} the typical loop induced electron coupling is of the order of
\begin{equation}
    g^{\rm loop}_{ae}\sim \left(\frac{\alpha}{\pi}\right)(m_{e}g_{a\gamma})\log(m_{e}g_{a\gamma}).
\end{equation}
Using this we can compare the decay rate into electrons due to this loop induced coupling to that directly into photons,
\begin{equation}
    \frac{\Gamma^{\rm loop}_{ae}}{\Gamma_{a\gamma}}\sim \left(\frac{\alpha}{\pi}\right)^2\log^2(m_{e}g_{a\gamma})\left(\frac{m_{e}}{m_{a}}\right)^2\ll 1 \quad\!\!{\rm for}\!\!\quad m_{e}<m_{a}.
\end{equation}
Therefore, whenever the decay is kinematically possible ($2m_{e}<m_{a}$), the rate of the loop induced decay into electrons is negligible compared to that directly into photons and Fig.~\ref{fig:bound_massive} is essentially unaffected.}

\section{Future perspectives}\label{sec:future}
As we have seen {\it Fermi}-LAT measurements of the diffuse background set very competitive constraints on the ALP parameter space.
How do we expect such limits to improve with future gamma-ray observations in the MeV range?

This energy window is foreseen to be explored by new instruments such as GAMMA-400~\cite{Topchiev:2017obi} and new missions in the MeV domain, such as eASTROGAM~\cite{DeAngelis:2017gra} and/or AMEGO~\cite{McEnery:2019tcm}.

The GAMMA-400 (Gamma Astronomical Multifunctional Modular Apparatus) telescope is designed to investigate the origin of cosmic gamma rays from 20 MeV up to 1 TeV (launch planned after 2026). With an analogous field of view as the {\it Fermi}-LAT, GAMMA-400 will improve significantly on angular (energy) resolution  above 10 (1) GeV. However, GAMMA-400 performances at 100 MeV will be comparable to {\it Fermi}-LAT, with a reduction of a factor $\sim$ 1.5 in effective area~\cite{Egorov:2020cmx}. 
As confirmed by comparing predictions for ALP constraints from detection of future SN explosions at different locations in the sky of Refs.~\cite{Egorov:2020cmx,Meyer:2016wrm}, we expect comparable performances of GAMMA-400 and the LAT also for the DSNALPB search. In particular, given the very similar angular resolution at low energies, we do not expect GAMMA-400 to improve the determination of the diffuse background by resolving more astrophysical sources.

The MeV domain (from hundreds of keV up to hundreds of MeV) remains strongly under-explored with sensitivity of past experiments being at least two orders of magnitude larger than what current telescopes can achieve.
Missions like eASTROGAM and AMEGO have been proposed to explore this energy window in the future but they have not been approved yet. 
We refer to the eASTROGAM instrument specification in~\cite{DeAngelis:2017gra} to predict what constraints on ALPs we would get from DSNALPB. We consider the pair-conversion domain ($0.03 - 3$ GeV) which is the most relevant for our purposes. If we perform a simple sensitivity estimate based on effective area performances, we can see that {\it Fermi}-LAT will still perform better than 10yr of eASTROGAM observations -- the effective area of {\it Fermi}-LAT being a factor of 10 larger than the one of eASTROGAM at 100 MeV. However, a 1yr observation with eASTROGAM will reach a factor of 2 better sensitivity to extragalactic objects than 10yr of {\it Fermi}-LAT data ($1.2 \times 10^{-12}$ erg/cm$^2$/s against $2.8\times 10^{-12}$ erg/cm$^2$/s at 100 MeV). This will boost our ability to resolve more sources currently adding up to the diffuse gamma-ray background, especially faint emitters such as misaligned AGN and star-forming galaxies, and to better constrain their contribution to the diffuse background. Ultimately, a better constraint of the diffuse background model components will improve our understanding of truly diffuse emission from, for example, the DSNALPB.

On the other hand, limits from the DNSALPB decay that we set by using data from COMPTEL, are expected to improve at least by a factor of 2 (but up to a factor of 40), depending on the achieved eASTROGAM effective area at the transition between the Compton and the pair-conversion domain.

\section{Conclusions}\label{sec:conclusions}
Applying the energy loss argument to  SNe sets stringent limits on a wide range of couplings and masses of ALPs. 
That said, huge amounts of energy could still be emitted into such novel channels.
Indeed, ALPs saturating the SN 1987A limit are emitted as copiously as neutrinos from SN cores. In this situation one expects not only a strong ALP burst from each SN, but also a large cosmic diffuse background flux from all past SNe, the DSNALPB~\cite{Raffelt:2011ft} in analogy to the diffuse SN neutrino background.  

In the minimal scenario in which ALPs couple only to photons, 
they would be produced in a SN core via the Primakoff process. 
 Then the 
 diffuse SN ALP flux would convert into gamma rays in the Galactic magnetic field. 
Using the recent measurement of the diffuse gamma-ray flux by the {\it Fermi}-LAT experiment, this allows us to place  a new bound on the photon-ALP coupling.  For $m_a \lesssim 10^{-10}$ eV it is slightly better than the current limit from CAST. While it is not as strong as the limit from SN 1987A, it provides a complementary check since it does not depend on a single event.
The bound significantly improves for scenarios in which ALPs couple also to nucleons, where  one would significantly enhance the ALP emissivity in a SN through the NN bremsstrahlung  process. Taking the ALP-nucleon coupling close to the limit from SN 1987A, the diffuse gamma-ray flux allows one to improve the bound on $g_{a\gamma}$ by two orders of magnitude (cf.~Fig.~\ref{fig:overview}) and by about three orders of magnitude if we use the constraints on the gamma-ray flux during the single event of SN 1987A. Both these observations also set stringent constraints of an explanation of X-ray observations of neutron stars~\cite{Buschmann:2019pfp} in terms of ALPs coupling to both photons and neutrons.

Massive ALPs (with $m_a \gtrsim 1$ keV) are also constrained from
 the diffuse gamma-ray flux produced by their decay into photons (see also~\cite{DeRocco:2019njg}).
Even for a pure photon coupling this provides the best limit in a mass range of roughly ($10-100$)~keV. Allowing for an additional nucleon coupling the sensitivity to the photon coupling could be significantly better, exceeding all other limits by several orders of magnitude, as can be seen from Fig.~\ref{fig:overview}.

Even if, as is the case for light ALPs, the SN 1987A bounds dominate over the one from the DSNALPB, it is important to stress that they have different systematic uncertainties and recognized or unrecognized loop holes. Therefore, to corner ALPs it is important to use as many independent approaches as possible.
 
Importantly, inclusion of different couplings (e.g. to nucleons or electrons) can dramatically change the sensitivity of the chosen approaches. 
Beyond that, considering different measurements may also allow us to determine individual couplings\footnote{For a possibility to measure the ALP photon and electron couplings using their differing energy spectra in IAXO~\cite{Irastorza:2011gs,Armengaud:2014gea} see~\cite{Jaeckel:2018mbn}.}, thereby learning more about the underlying fundamental structures.
For example, in the optimistic scenario that we can observe gamma rays from the DSNALPB (or a future SN) and have a positive signal in IAXO~\cite{Irastorza:2011gs,Armengaud:2014gea} we have a chance to determine both the photon as well as the nucleon coupling. Similarly the combination with a dark matter experiment sensitive to either the photon~\cite{Sikivie:1983ip,Horns:2012jf,Jaeckel:2013sqa,Shokair:2014rna,Kahn:2016aff,Petrakou:2017epq,Majorovits:2017ppy,McAllister:2017lkb,
Alesini:2019nzq,Melcon:2018dba,Droster:2019fur, Alesini:2019ajt}  or the nucleon coupling~\cite{Budker:2013hfa}  would allow such a measurement.

Further improvements may be achieved by performing a dedicated analysis of the anisotropy in the diffuse gamma-ray spectrum measured by {\it Fermi}-LAT. Indeed, one expects that in the case of ALPs, the morphology of the Galactic magnetic field would imprint special features in the produced gamma-ray flux. 
Moreover, we have discussed how future gamma-ray experiments in the MeV range have potential to improve the sensitivity.
Therefore, we hope that this approach of searching ALPs will continue in the future, reaching a level of sophistication similar
to the one currently adopted in WIMP dark matter searches.

\section*{Acknowledgments}
The work of P.C. and 
A.M. is partially supported by the Italian Istituto Nazionale di Fisica Nucleare (INFN) through the ``Theoretical Astroparticle Physics'' project
and by the research grant number 2017W4HA7S
``NAT-NET: Neutrino and Astroparticle Theory Network'' under the program
PRIN 2017 funded by the Italian Ministero dell'Universit\`a e della
Ricerca (MUR).

\bibliographystyle{utphys}
\bibliography{references}

\end{document}